\documentstyle[twocolumn,eqsecnum,aps]{revtex}
\input boxedeps.tex
\SetRokickiEPSFSpecial  
\HideDisplacementBoxes

\newcommand{\D}{\overline{D}}
\newcommand{\da}{\dagger}  
\newcommand{\be}{\begin{equation}}
\newcommand{\eq}{\end{equation}}
   
\newcommand{\Tr}{{\rm \, Tr \!}}    
\newcommand{\dm}{{\cal M}}          

\begin{document}
\draft
\preprint{\parbox{1.5in}{FAU--TP3--98/20\\ 
                         CERN--TH/98--347\\ 
                       {\tt hep-th/9810236}}}
\title{Glueball calculations in large-$N_c$ gauge theory}
\author{S. Dalley}
\address{Theory Division, CERN, CH--1211 Geneva 23, Switzerland}
\author{B. van de Sande}
\address{Institut F\"ur Theoretische Physik III,
Staudstra{\ss}e 7, D--91058 Erlangen, Germany}
\maketitle
\begin{abstract}
We use the light-front Hamiltonian of
transverse lattice gauge theory to compute from first principles  
the glueball spectrum and light-front wavefunctions
in the leading order of the $1/N_{c}$ colour expansion. 
We find $0^{++}$, $2^{++}$, and $1^{+-}$ glueballs having masses 
consistent with $N_{c}=3$ data available from 
Euclidean lattice path integral methods.
The wavefunctions exhibit a light-front constituent gluon structure.

\end{abstract}

\section{Introduction}

There is growing experimental evidence that glueballs,  
boundstates of gluons in the $SU(3)$ gauge theory
Quantum Chromodynamics (QCD), have
been discovered in the mass range $1.5-1.7$~GeV
\cite{close1,ibm,lee}. 
But the confinement feature of QCD --- 
interactions grow stronger 
as the energy of a process decreases --- 
complicates any first principles calculation
of the boundstate problem.
To date, the most successful boundstate calculations 
use a Euclidean spacetime lattice and simulate 
the path integral by Monte Carlo methods.
These difficult calculations are now 
roughly consistent with the experimental 
signatures of glueballs \cite{ibm,ukqcd}, 
although much detail remains to be clarified.
Therefore, it is important to have some independent method of calculating
the properties of glueballs
from first principles in gauge theory. 
In this letter, we present our first results on this problem 
using an effective light-front Hamiltonian quantisation
(canonical quantisation on a null-plane in spacetime). 
This is the transverse lattice method, suggested
originally by Bardeen and Pearson \cite{bard1,bard2}, which we have
developed to the extent that quantitative
calculations are now feasible \cite{dv1,dv2}.

Although this work is ostensibly about glueballs, 
we have a more general 
motivation for developing the light-front Hamiltonian
formulation of gauge theory. A light-front Hamiltonian
has Lorentz-frame-independent wavefunctions. Together with
the simplicity of the light-front vacuum, this leads to a field-theoretic
realisation of the parton model for hadrons
on which so much understanding is based.
The rich phenomenology in hadronic and nuclear physics
that would follow from knowledge of the 
light-front wavefunctions is surveyed in the lectures of Brodsky~\cite{stan}.
We will use the glueball problem in QCD, which is especially difficult 
computationally, as a quantitative test of our light-front formalism.

A detailed account of our methods and various quantitative
tests that we have performed, mostly in $2+1$ dimensions,
can be found in Refs.~\cite{dv1,dv2} but we will briefly review the 
salient points below. 
We will work in the leading order of the $1/N_c$-expansion of $SU(N_c)$ 
gauge theory, which omits the $1/N_{c}^{2}$-suppressed glue configurations.
It also removes ``sea'' quarks from our theory, an approximation
also used in the Euclidean lattice calculations.
We work on a coarse transverse lattice, using an effective potential tuned
to minimize discretisation errors. 
The groundstate ${\cal J}^{\cal PC} =0^{++}$ glueball mass is 
${\cal M} = 3.3 \pm 0.2
\sqrt{\sigma}$, where $\sigma \approx 0.1936 {\rm GeV}^2$ 
is the string tension, and is consistent with $SU(2)$ and
$SU(3)$ results available from the Euclidean lattice. Components of
the $2^{++}$ and $1^{+-}$ glueballs which are behaving covariantly 
are also consistent. The glueball wavefunctions that we obtain are new.

\section{Transverse Lattice Formulation}

In $3+1$ spacetime dimensions we use a square lattice of spacing $a$
in the `transverse' directions ${\bf x}=\{x^1,x^2\}$, 
and a continuum in the $\{x^0,x^3\}$ directions.
Our action is 
\begin{eqnarray}
S &  = & \int dx^0 dx^3 \sum_{{\bf x}}  \left(
  \Tr\left\{ \D_{\alpha} M_r({\bf x}) \left(\D^{\alpha} 
M_r({\bf x})\right)^{\da}
\right\} \right. \nonumber \\
&& \left. \hspace{15mm} 
-  {1 \over 2G^2} \Tr\left\{F_{\alpha\beta}F^{\alpha\beta}\right\}  
- V_{{\bf x}}[M]\right) \;  ,
\label{lag}
\end{eqnarray}
where $r,s\in \{1,2\}$; $\alpha,\beta\in\{0,3\}$; and
\begin{equation}
\D_{\alpha} M_r({\bf x})  
         =   \left(\partial_{\alpha} +i A_{\alpha} ({\bf x})\right)
        M_r({\bf x})  
        -  i M_r({\bf x})   A_{\alpha}({{\bf x}+a \hat{\bf r}})\; .
\end{equation}
$A_{\alpha}$({\bf x}) is the gauge potential at transverse site 
${\bf x}$, while
$M_r({\bf x}) \in GL(N_c,C)$ is a link variable from ${\bf x}$ to 
${\bf x} + a \hat{\bf r}$ which carries colour flux between sites. 
Our use of  non-compact variables at finite $a$
is the basis of the colour-dielectric formulation of lattice gauge theory
\cite{hans}. The action (\ref{lag}) is gauge invariant at each transverse 
lattice site under
\begin{eqnarray}
        A_{\alpha}({\bf x}) & \to & U({\bf x}) A_{\alpha}({\bf x}) 
        U^{\da}({\bf x}) + {\rm i} \left(\partial_{\alpha} U({\bf x})\right) 
        U^{\da}({\bf x})  \\
        M_r({\bf x}) &  \to & U({\bf x}) M_r({\bf x})  
        U^{\da}({\bf x} + a\hat{\bf r})  
\end{eqnarray}
for $U \in SU(N_c)$.
The dimensionful coupling $G^2(a)$ is such that 
$a^{2} G^2 \to g^{2}$ in the classical continuum limit $a \to 0$, 
where $g^{2}$ is the usual continuum gauge coupling. 
We include a gauge-invariant effective potential
$V_{{\bf x}}[M]$ consisting of Wilson flux loops on the transverse
lattice. The exact form we use is given below.  Part of its job is to
ensure that $M$ is forced to be a unitary matrix as $a \to 0$, so that 
continuum QCD is recovered \`{a} la Wilson \cite{wilson}.
More generally, we choose it to minimize violations of Lorentz covariance
at finite $a$. The idea is that we tune the couplings of $V$ along
an  approximation to the Lorentz covariant scaling trajectory which leads to 
continuum QCD as $a \to 0$.
Then for low energy observables, such as boundstate masses,
one can work at finite $a$.

We perform light-front quantisation, treating $x^{+} = (x^0 + x^3)/\sqrt{2}$
as time, and pick the light-front gauge $A_{-}= (A_{0}-A_{3})/\sqrt{2} = 0$.
In this gauge $A_{+} = (A_{0}+ A_{3})/\sqrt{2}$ is a constrained field
which we eliminate by solving its equation of motion classically. 
Working in the leading $1/N_c$ approximation has some simplifying 
consequences. In the transverse rest frame ${\bf P}=0$, Eguchi-Kawai
reduction occurs at large $a$, so that the argument ${\bf x}$
may be dropped \cite{dv1}. This results in a light-front Hamiltonian of the
form
\begin{eqnarray}
 P^-  & = & \int dx^- 
 - {G^2 \over 4} \Tr\left\{  J^{+} \frac{1}{\partial_{-}^{2}} J^{+} \right\} 
\nonumber \\
&& -{\beta \over N_ca^{2}} \Tr\left\{ M_{2}^{\da} M_{1}^{\da}
M_{2} M_{1} 
 +  M_{1} M_{2} M_{1}^{\da} M_{2}^{\da}
\right\} \nonumber \\
&& + \mu^2   \sum_r  
 \Tr\left\{M_r M_r^{\da}\right\} 
 + {\lambda_3 \over a^{2} N_{c}^{2}}  \sum_r   
\left( \Tr\left\{ M_r M_r^{\da} \right\} \right)^2 
\nonumber \\ 
&& +
 {\lambda_2 \over a^{2} N_c}\sum_r  
\Tr\left\{ M_r M_r
M_r^{\da} M_r^{\da} \right\} \nonumber \\
&& + {\lambda_1 \over a^{2} N_c} \sum_r  
\Tr\left\{ M_r M_r^{\da}
M_r M_r^{\da} \right\} 
\nonumber \\
&& +  {\lambda_4 \over a^{2} N_c}  
\sum_{\sigma=\pm 2, \sigma^\prime = \pm 1}
        \Tr\left\{ 
M_\sigma^{\da} M_\sigma M_{\sigma^\prime}^{\da} M_{\sigma^\prime} \right\} 
        \nonumber\\
&& +  {4 \lambda_5 \over a^{2} N_{c}^{2}} 
\Tr\left\{ M_1 M_1^{\da} \right\}\Tr\left\{ M_2 M_2^{\da} \right\} \ ,
        \label{redham}\\
J^{+} &=&  {\rm i} ( M_r \stackrel{\leftrightarrow}{\partial}_{-} 
M_r^{\da}  + M_r^{\da} 
\stackrel{\leftrightarrow}{\partial}_{-} M_r)\; .
\end{eqnarray}
%
This Hamiltonian includes all  Eguchi-Kawai-reduced Wilson loops on the
transverse lattice up to fourth order in $M$;
this is our
approximation to the effective potential $V$. 
Generalisation to higher orders is straightforward.
The Hamiltonian is diagonalised in a parton Fock space of
$M_{r}$, at fixed total momentum $P^+$, built from its Fourier modes 
at $x^+ = 0$
\begin{eqnarray}
M_r &  =  & 
        \frac{1}{\sqrt{4 \pi }} \int_{0}^{\infty} {dk^+ \over {\sqrt{ k^+}}}
        \left( a_{-r}\, e^{ -i k^+ x^-}  +   a^{\da}_r\,
 e^{ i k^+ x^-} \right) \; .
\end{eqnarray}
However, only closed flux loops, invariant under
residual $x^-$-independent gauge transformations $U$, have finite
energy. These loops propagate without splitting or joining at leading
$1/N_c$ order.

Even with a finite lattice spacing $a$, the Hamiltonian (\ref{redham}) 
is not yet
completely regulated. We impose periodic boundary conditions in $x^-$
and a cut-off on the maximum number of link partons $a_{r}$ in the Fock space. 
For each
choice of couplings in the Hamiltonian, we remove these latter
cut-offs by extrapolation, following a continuum improvement
procedure.

\section{Tuning the Effective Potential}

To test for Lorentz covariance, we measure the dispersion relations
of glueball boundstates
as a function of couplings in $V$. This requires
us to work at ${\bf P} \neq 0$, which can be achieved by adding appropriate
phase factors to matrix elements of $P^-$ above \cite{dv2}. 
Using $G^2 N_c$ to set the
dimensionful scale, the dispersion relation
of each glueball can be written
\begin{eqnarray}
      2P^+ P^- & = &  G^2 N_c \left( \dm^{2}_{0} + \dm_{1}^{2}\, a^2
      |{\bf P}|^2  \right. \nonumber \\
&& \left. \hspace{10mm} + 2 \overline{\dm}_{1}^{2} \, a^2 P^1 P^2  
+ O(a^4 |{\bf P}|^4) \right) \label{latshell} \; .
\end{eqnarray}
For each glueball in the light spectrum and for fixed $\mu^2$,
we tune the couplings $\lambda_i$ and $\beta$ so that,
as far as possible, 
\begin{eqnarray}
a^2 G^2 N_c \dm_{1}^{2}  \equiv c_{\rm on}^2 & = & 1 \\
a^2 G^2 N_c (\dm_{1}^{2}+\overline{\dm}_{1}^{2}) \equiv 
c_{\rm off}^2 & = & 1  \; . 
\end{eqnarray}
$c_{\rm on}$ is the speed of light in direction ${\bf x}=(1,0)$;
$c_{\rm off}$ is the speed of light in the direction ${\bf x}=(1,1)$.
In practice, we apply a $\chi^2$-test over a range of light glueballs to
optimize isotropy of the speed of light.
In this way we search for a Lorentz covariant
scaling trajectory parameterized by $\mu^2$,
which may later be related to $a$.

It is necessary to know the dimensionless combination
$a^2 G^2 N_c$ for this procedure. This can be
deduced from two measurements of the string tension $\sigma$. 
The first measures
the mass squared of winding strings in the transverse lattice directions 
${\bf x}$, fit to $n^2 a^2 \sigma^{2}_{T}$ for winding number $n$. 
The second measures the
heavy quark potential in the continuous $x^3$ direction \cite{burkardt}, 
fit to
$\sigma_L R$ for separation $R$. Demanding $\sigma_T=\sigma_L \equiv \sigma$
and expressing eigenvalues in units of
$G^2 N_c$, we may eliminate $\sigma$ and deduce 
$a^2 G^2 N_c$. We may also deduce $G^2 N_c$, and hence all dimensionful 
quantities, in units of $\sigma$.

To further help optimize lattice discretisation errors, we include in
the $\chi^2$ test a measure of deviations from rotational invariance
of the heavy-quark potential, including deviations from
$\sigma_T = \sigma_L$ and deviations from Lorentz multiplet
structure in the glueball spectrum. $\sigma_L/G^2 N_c$
is the only basic quantity which is  poorly determined by our 
current approximations, and our results were particularly sensitive to the 
variance assignment in the
$\chi^2$ test for this datum
(though still compatible with the systematic error estimate
we make below). 

According to the colour-dielectric
picture, large $\mu$ will correspond to large $a$ where discretisation
errors grow,
while small $\mu$ corresponds to intermediate spacing $a$. 
Sufficiently small $a$ produces
a non-trivial light-front vacuum, signalled by the appearance of
tachyons; it would be difficult to 
construct an effective potential in this regime. 
Instead, we use an effective potential 
at intermediate $a$ on a trivial light-front vacuum.
In $2+1$ dimensions the above procedure
accurately identified a Lorentz covariant
scaling trajectory at intermediate $a$ \cite{dv2}.
The 
results we present below are based on a similar trajectory
which we have found in $3+1$ dimensions, 
though further work is needed to reduce systematic errors.

\section{Results}

In Table~\ref{table1} we show results for various components of the lightest
glueballs at a transverse lattice spacing $a = 1.1 /\sqrt{\sigma} 
\approx 0.49\,{\rm fm}$, at 
the overall $\chi^2$ minimum for this $a$. 
In order to establish the scaling trajectory,
we have performed calculations at other lattice spacings:
At smaller lattice spacings tachyons appear, indicating
breakdown of the trivial vacuum regime. 
For larger lattice spacings
the couplings evolve gradually, and the minimum $\chi^2$ value grows.
For the groundstate $0^{++}$ we find approximate scaling and maintenance of
Lorentz covariance 
 $(a/\sqrt{\sigma},{\cal M}/\sqrt{\sigma},c)
= \{(1.1,3.3,1.13),(1.2,3.1,1.18),(1.3,3.5,1.19) \}$.
The errors shown on
glueball masses in Table~\ref{table1} are estimates of the error in 
removing the periodic
boundary condition on $x^-$ and the cut-off on the maximum number 
link variables $M_{r}$ in a wavefunction; these errors are under
control. 
Only for the $0^{++}$ mass are we confident enough to estimate the
 finite-$a$ systematic error.  
The Lorentz covariant scaling trajectory will need to be determined 
more accurately before similar statements can be made for the higher states. 
In $2+1$ dimensions, we found that 
when the deduced speed of light becomes isotropic to within
about 10\%, the corresponding glueball mass is also accurate at this 
level; this also seems to be the case in 3+1 dimensions.

\begin{table}
\centering\[
\renewcommand{\arraystretch}{1.25}
\begin{array}{cc|cc|c} 
\displaystyle {\cal J}^{\cal P C} &|{\cal J}_z|^{{\cal P}_1} &  c_{\rm on} &
c_{\rm off}  & \displaystyle{{\cal M} \over \sqrt{\sigma}} \\ \hline \hline
0^{++} & 0^{+}  & 1.13 & 1.13 & 3.3(1) \\ \hline
2^{++} & 2^{+}  & 0.85 & 0.86 &  4.4(1)   \\ 
       & 2^{-}  & {\rm Im} & - & 5.9(2)  \\ 
       & 1^{\pm}& 0.89,- & 0.92, {\rm Im} & 5.4(2) \\ 
       & 0^{+}  & 0.84 & 0.83 & 4.5(1)   \\
\hline 
1^{+-} & 1^{\pm}&  0.75, 0.88 & 
                   0.74, 0.89 & 5.0(1)  \\ 
       & 0^{-}  &  0.99 & 0.99 & 5.9(2)    
\end{array}\]
\caption{The glueball multiplet components 
showing masses ${\cal M}^2 = 2P^+ P^- ({\bf P} = 0)$ and 
$c$ from  the dispersion relation (\protect\ref{latshell}). 
${\cal P}_1$ is the symmetry under reflections $x^1\to-x^1$;
``Im'' indicates $c^2 < 0$ and ``$-$'' indicates
the quantity has not been measured.
The ${\cal J}_z = \pm 1$ states are exactly degenerate. 
\label{table1}}
\end{table}

\begin{figure}
\centering
$\frac{M}{\sqrt{\sigma}}$\BoxedEPSF{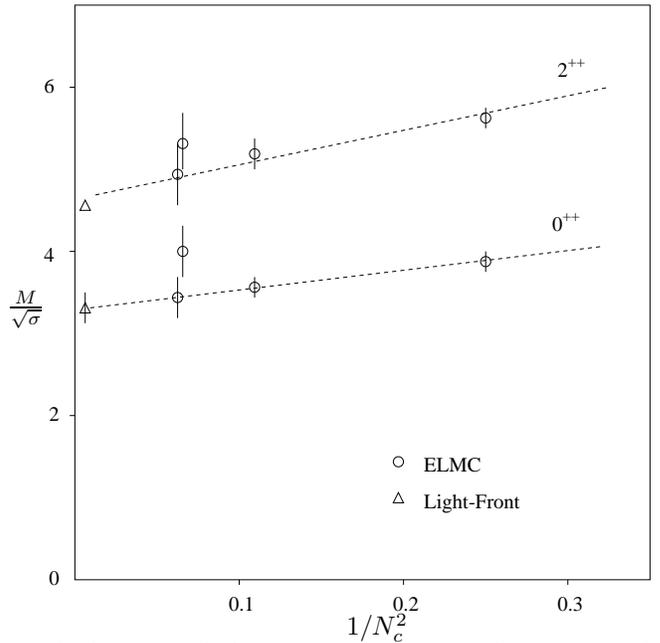 scaled 430}\\
\hspace{0.5in}$1/N_{c}^{2}$
\caption{The variation of glueball masses with $N_c$. 
ELMC predictions are continuum ones for $N_c=2,3$, and fixed lattice 
spacings (higher points are $\beta=11.1$ and lower are $\beta = 10.9$) 
for $N_c = 4$ \protect\cite{mike}. $N_c = \infty$ light-front predictions are
from components behaving covariantly in Table~\ref{table1}, including
the total error estimate for the $0^{++}$. The dotted lines 
are to guide the eye, and correspond to the expectation of leading 
linear dependence
on $1/N_{c}^2$.
\label{fig1}}
\end{figure}

We now compare our results with spectra from conventional
Euclidean lattice Monte Carlo (ELMC) simulations.
See Fig.~\ref{fig1}. 
(Recently, large $N_c$ glueball mass ratios
have been calculated using a conjectured
duality with supergravity~\protect\cite{super}; 
we will not discuss these results since 
their errors are not 
quantified.)

\noindent 
\underline{$0^{++}$}\\
Based on violations of Lorentz covariance
and variation of ${\cal M}$ along the
scaling trajectory, we include an estimate of 
systematic error for this state,
${\cal M} = 3.3\pm 0.2 \sqrt{\sigma}$.  
A continuum extrapolation
of $SU(3)$ ELMC data has been made in Ref.~\cite{close2}, 
${\cal M} = 3.65 (15) \sqrt{\sigma}$ (statistical error).
%
%
%
Furthermore, measurements
for $SU(2)$ and $SU(4)$ \cite{mike} indicate a slight downward drift with 
decreasing $1/N_{c}^{2}$.

\noindent \underline{$2^{++}$}\\
Not all components of this state are behaving covariantly. The
${\cal P}_1$-odd combinations of ${\cal J}_z = \pm 1$ and 
${\cal J}_z = \pm 2$ have large
lattice artifacts and cannot be trusted.  As a result, we are unsure of
the correct parity assignment for the ${\cal J}_z = \pm 1$ but assume here
that it belongs to the $2^{++}$ multiplet. 
The ${\cal P}_1$-even ${\cal J}_{z}= 0$ and ${\cal J}_{z} = \pm 2$ combination
are nearly degenerate and isotropic, 
pointing to a tensor mass near 
${\cal M} \approx 4.5 \sqrt{\sigma}$.  The $SU(3)$ 
ELMC extrapolation of Ref.~\cite{close2} gives
$5.15(23)$, consistent with more recent coarse anisotropic ELMC
simulations \cite{star}.
Again, there is  a slight drift with $1/N_{c}^2$. 

\noindent \underline{$1^{+-}$}\\
Here the ${\cal P}_1$-odd components fair better. The
${\cal J}_z = 0$ component of the lightest spin 1 
is behaving covariantly,
and points to ${\cal M} \approx 5.9 \sqrt{\sigma}$. 
The ${\cal J}_{z}=  \pm 1$ component is
split away as a consequence of its anisotropy. 
An $SU(3)$ value ${\cal M} = 6.6(6) \sqrt{\sigma}$ is given in 
Ref.~\cite{ukqcd}, while the more accurate 
measurement of Ref.~\cite{star}
%
%
implies ${\cal M}=6.30(4)\sqrt{\sigma}$.

\noindent \underline{$0^{-+}$}\\
Our lightest candidate for this state appears around  $7.4 \sqrt{\sigma}$
though we have not measured its covariance. 
$SU(3)$ ELMC work suggests a lighter mass~\cite{ukqcd}. 
Barring an unusual large-$N_c$ extrapolation, it seems likely
that this state (which posed problems for early ELMC studies also)
is poorly described with our current approximations.

\begin{figure}
\centering
$G_d(x)$\BoxedEPSF{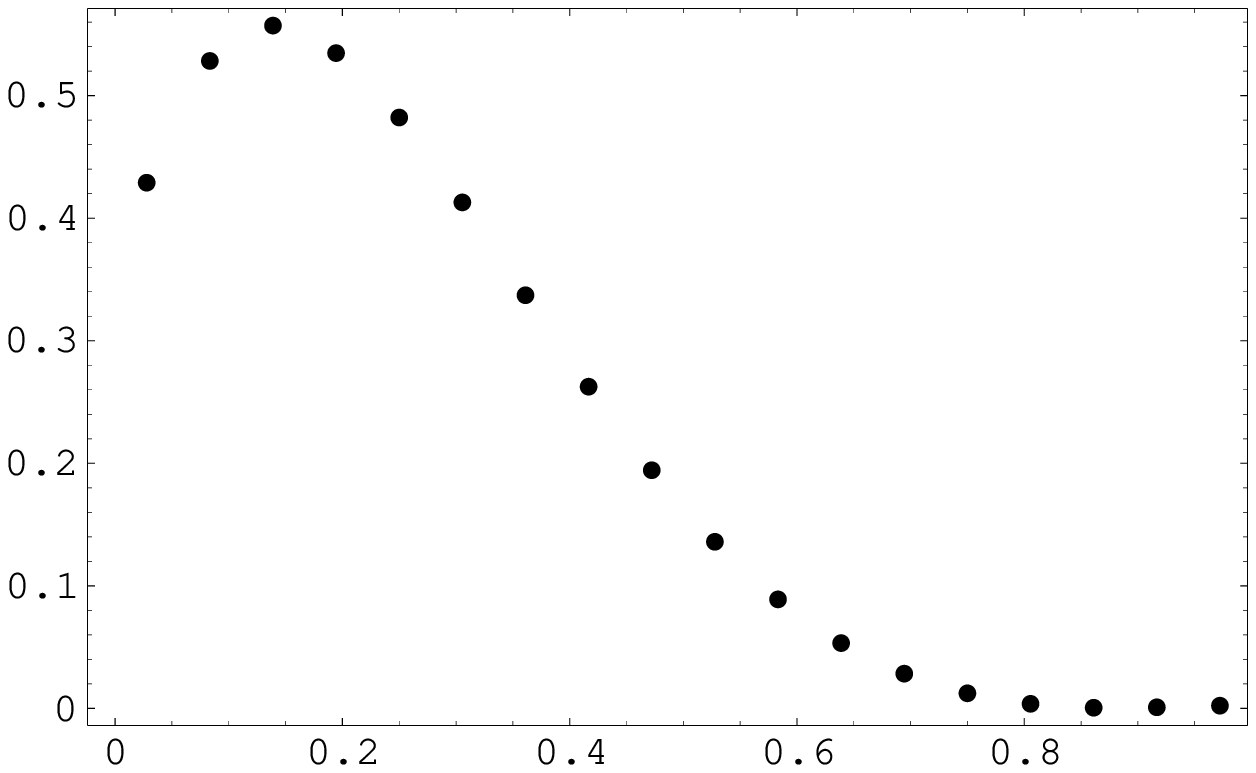 scaled 600}\\
\hspace{0.5in}$x$
\caption{The $0^{++}$ momentum distribution function.
\label{fig2}}
\end{figure}

As an example of what may be deduced from wavefunctions, 
in Figure~\ref{fig2} we 
plot the distribution of longitudinal momentum fraction 
$x = k^+ / P^+$ in the $0^{++}$ glueball at $a = 1.1/\sqrt{\sigma}$. 
$G_d(x)$ is the 
probability distribution for a link parton $a_r (k^+)$
and becomes the usual gluon distribution function in the continuum
limit $a \to 0$ \cite{dv2}.
At finite $a$, $G_d$ may be interpreted as a light-front
constituent gluon distribution that evolves with $a$. 
We find that our 
$0^{++}$ light-front wavefunction consists mostly of four such 
constituents; 
it does {\em not} obey a gluonium picture (two constituent gluons).

\section{Conclusions}

We have measured glueball masses in QCD,  using the $1/N_c$ expansion
and different quantisation, field variables, regulators, and 
gauge fixing from the traditional 
Euclidean lattice path integral simulations. The results
from both approaches are consistent for QCD without quarks,
indicating a $0^{++}$
glueball near $1.6$ GeV, and $1/N_{c}^2$ corrections 
appear to be small for the $0^{++}$ and $2^{++}$ states. 
(In this latter regard, it would be interesting 
to see better data at $SU(4)$.) 
We found that light-front wavefunctions exhibit a constituent gluon
structure, although not of the form one might naively guess.
In order to improve the accuracy of the Lorentz covariant scaling 
trajectory,
future work includes choosing other observables to set the scale 
and incorporating higher order effective interactions.
Finally, we note that the re-introduction of quarks will be necessary
for a detailed comparison with experimental glueball candidates, which 
mix with nearby scalar quarkonia \cite{close1,lee}.

\acknowledgements{
The work of SD is supported by CERN. We thank M. Teper for helpful
interactions.  A portion of the computations were performed at
Argonne's Center for Computational Science and Technology.
}

\end{document}